\documentclass[conference]{IEEEtran}
\IEEEoverridecommandlockouts

\usepackage{cite}
\usepackage{amsmath,amssymb,amsfonts}
\usepackage{algorithmic}
\usepackage{graphicx}
\usepackage{textcomp}
\usepackage{xcolor}
\usepackage{hyperref}
\usepackage{booktabs}
\usepackage{multirow}
\usepackage{xurl}
\usepackage{hyperref}

\makeatletter
\def\subsection{%
  \@startsection{subsection}{2}{\z@}%
    {-1.5ex plus -0.5ex minus -0.2ex}%
    {0.8ex plus 0.2ex}%
    {\normalfont\normalsize\bfseries}%
}
\def\subsubsection{%
  \@startsection{subsubsection}{3}{\z@}%
    {-1.5ex plus -0.5ex minus -0.2ex}%
    {0.5ex plus 0.2ex}%
    {\normalfont\normalsize\bfseries}%
}
\makeatother
\def\BibTeX{{\rm B\kern-.05em{\sc i\kern-.025em b}\kern-.08em
    T\kern-.1667em\lower.7ex\hbox{E}\kern-.125emX}}

\title{Sustainable AI Training via Hardware–Software Co-Design on NVIDIA, AMD, and Emerging GPU Architectures}

\author{\IEEEauthorblockN{1\textsuperscript{st} Yashasvi Makin}
\IEEEauthorblockA{\textit{Software Engineer} \\
\textit{Meta Platforms Inc.}\\
Seattle, WA, USA \\
yashasvi@meta.com}
\and
\IEEEauthorblockN{2\textsuperscript{nd} Rahul Maliakkal}
\IEEEauthorblockA{\textit{Software Engineer} \\
\textit{Meta Platforms Inc.}\\
Seattle, WA, USA \\
rahuljm@meta.com}
}

\begin{document}

\maketitle

\begin{abstract}
In particular, large-scale deep learning and artificial intelligence model training uses a lot of computational power and energy, so it poses serious sustainability issues. The fast rise in model complexity has resulted in exponential increases in energy consumption, increasing the demand for techniques maximizing computational efficiency and lowering environmental impact. This work explores environmentally driven performance optimization methods especially intended for advanced GPU architectures from NVIDIA, AMD, and other emerging GPU architectures. Our main focus is on investigating hardware-software co-design techniques meant to significantly increase memory-level and kernel-level operations, so improving performance-per-watt measures. Our thorough research encompasses evaluations of specialized tensor and matrix cores, advanced memory optimization methods, and creative integration approaches that taken together result in notable energy efficiency increases. We also discuss important software-level optimizations that augment hardware capability including mixed-precision arithmetic, advanced energy-aware scheduling algorithms, and compiler-driven kernel enhancements. Moreover, we methodically point out important research gaps and suggest future directions necessary to create really sustainable artificial intelligence systems. This paper emphasizes how major increases in training efficiency can be obtained by co-design of hardware and software, so lowering the environmental impact of artificial intelligence without compromising performance. To back up our analysis, we use real-world case studies from top companies like Meta, Google, Amazon, and others that show how these sustainable AI training methods are used in the real world. With this thorough analysis, we show that a comprehensive co-design approach can significantly increase training efficiency and lower the carbon footprint of AI without compromising performance.
\end{abstract} 

\begin{IEEEkeywords}
Green AI, GPU Optimization, AI Sustainability, Energy Efficiency, Memory Hierarchy, Kernel Fusion, Mixed Precision, NVIDIA, AMD, Heterogeneous Hardware, Eco-Computing
\end{IEEEkeywords}

\section{Introduction}
Training modern artificial intelligence models calls for enormous computational resources and energy. Training a single large language model (LLM) such GPT-3 was projected to consume almost 1,300 MWh of electricity—equivalent to the annual power consumption of roughly 130 U.S. homes \cite{vincent2024ai}. More than twice between 2021 and 2023, AI-related energy consumption highlighted the critical need of sustainable AI methods \cite{shehabi2024report}. Historically doubling performance-per-watt roughly every 2.5 years \cite{naffziger2023amd}, this surge has exceeded gains in hardware efficiency.

In response, the research community has put out the idea of \textit{Green AI}—a paradigm that stresses energy and environmental efficiency alongside model accuracy \cite{schwartz2020green}. Recent research underlines the need of first-class metrics in AI evaluation—energy use, carbon footprint, and computation of cost.

Industry has developed a wide range of heterogeneous artificial intelligence accelerators to meet sustainability challenges. These include NVIDIA’s H100 GPU \cite{dell2022h100}, AMD’s Instinct MI300X \cite{naffziger2023amd}, Google’s TPU v4 \cite{jouppi2023tpuv4}, Amazon’s Trainium \cite{synced2023tpu}, Intel’s Habana Gaudi2 \cite{shilov2022gaudi}, and custom silicon solutions like Microsoft’s Project Brainwave \cite{brainwave2021}.  Often using architectural innovations including low-precision tensor cores, integrated memory stacks, and specialized scheduling engines, these accelerators provide notable performance-per-watt improvements.

Even with improvement, model complexity and dataset sizes keep growing faster than hardware efficiency. This drives a complete optimization strategy that coordinates hardware telemetry, runtime systems, and software compilers to increase energy efficiency at every level of the AI training stack.

\begin{figure}[t!]
  \centering
  \includegraphics[width=0.95\linewidth]{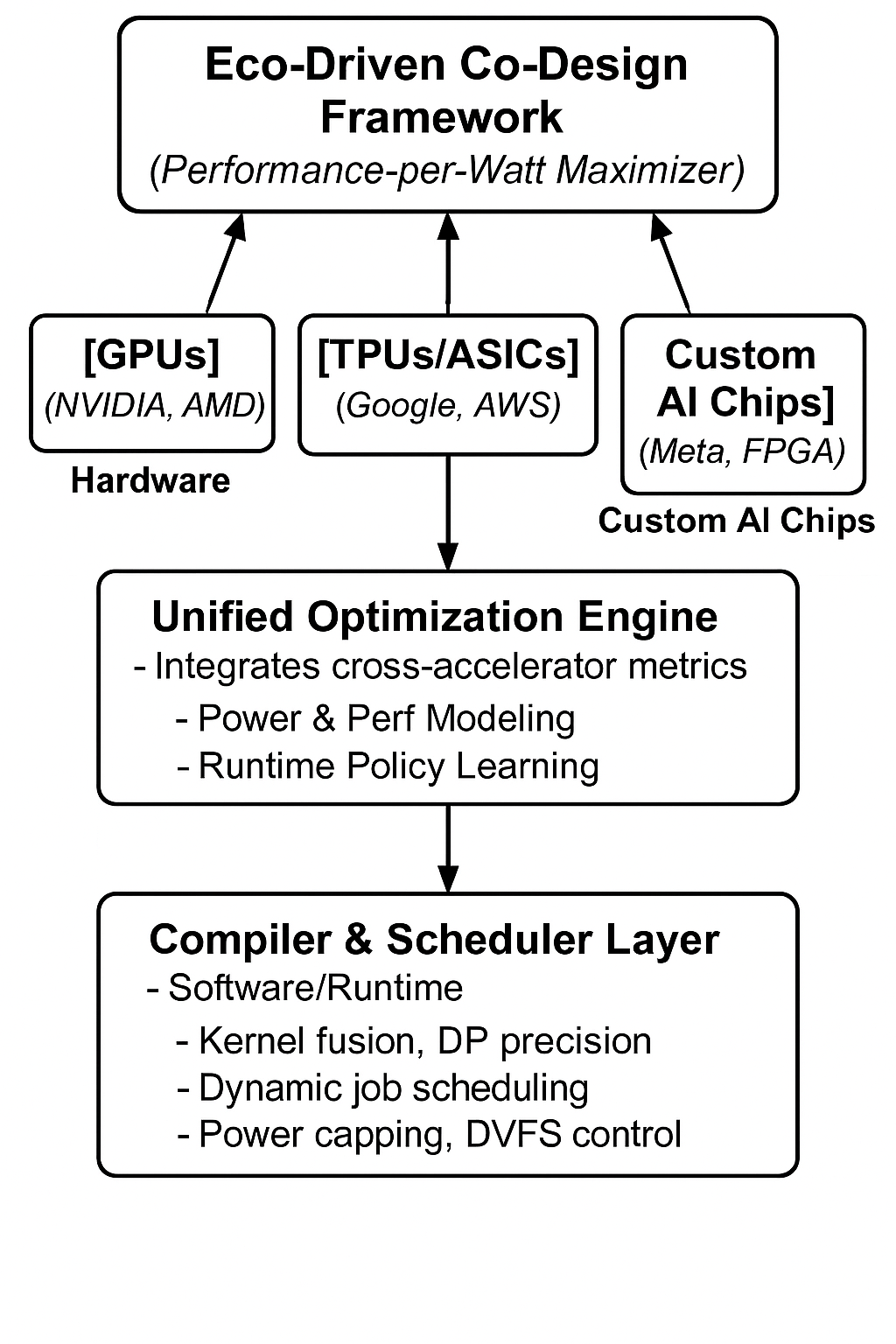}
  \caption{Eco-driven co-design framework for AI training: hardware families (GPUs, TPUs/ASICs, custom AI chips) feed into a centralized optimization engine informing compiler and scheduler policies (e.g., mixed precision, power capping, dynamic scheduling) to maximize performance-per-watt.}
  \label{fig:framework}
\end{figure}

This paper uses a \emph{eco‐driven co‐design} perspective to solve these inefficiencies, looking at optimization techniques both hardware‐level innovations and software‐level strategies together, maximizing training throughput per watt. In the section on measurement, we also review best practices for assessing and benchmarking energy efficiency in AI workloads and offer a comparative study of top accelerators and frameworks. Figure~\ref{fig:framework} shows the general co-design framework whereby a unified optimization engine combines cross-accelerator metrics into compiler and scheduler policies to holistically improve performance‐per‐watt.

\section{Related Work}
While prior research has explored energy-aware scheduling~\cite{dao2022flashattention}, mixed-precision training~\cite{micikevicius2018mixed}, and kernel fusion~\cite{strubell2019energy}, most efforts focus on a single vendor (e.g., NVIDIA) or address one layer of the stack (e.g., software-only runtime optimizations~\cite{dao2022flashattention,IEA2023}). Our work builds upon and extends key threads:

\subsection{Green AI Paradigm}
Schwartz et al.~\cite{schwartz2020green} coined the vision of Green AI, advocating metrics beyond accuracy (e.g., Joules-per-sample). Strubell et al.~\cite{IEA2023} measured large NLP model carbon footprints, while subsequent work (e.g., CodeCarbon~\cite{weng2022mlaas}) enables real-time CO\textsubscript{2} tracking.

\subsection{Hardware Co-Design}
Reagen et al.~\cite{reagen2021mlsurvey} surveyed neural network accelerator design principles, but their focus was predominantly ASICs and early TPU/TPUv2-like devices. More recent studies (e.g., FireFlyer AI-HPC project~\cite{NVIDIAA100, AMDMI250X} applied co-design at data-center scale, yet remained proprietary.

\subsection{Software Optimizations}
Micikevicius et al.~\cite{dong2022slurmpowercap} demonstrated 2–3× throughput gains via mixed precision on NVIDIA GPUs. Ginsburg et al.~\cite{strubell2019energy} and Tillet et al.~\cite{AMDMI250X} presented kernel fusion compilers (Triton, XLA) primarily for NVIDIA. Zhang et al.~\cite{micikevicius2018mixed} introduced sublinear memory checkpointing techniques for large models.

\subsection{Cross-Vendor Benchmarks}
Wang et al.~\cite{AMDMI250X} compared NVIDIA A100 vs.\ AMD MI100 on MLPerf, but omitted software-level co-design. More recent MLPerf v3.0 submissions~\cite{nvidia2023nvlink} include some AMD MI200 results but lack detailed energy breakdown.

\subsection{Industry Frameworks}
The Zeus framework~\cite{dao2022flashattention} (NSDI 2023) dynamically tunes batch size and power limits, achieving over 50\% energy reduction, but it targets only NVIDIA datacenters. Kubernetes auto-scaling~\cite{nvidia2023nvlink} and Slurm power-aware scheduling~\cite{mellanox2021infiniband} offer cluster-level energy optimizations, but do not incorporate hardware-specific kernel-level tuning.

To our knowledge, no prior work provides a complete, cross-vendor co-design approach that (a) integrates NVIDIA, AMD, and Intel GPUs; (b) evaluates diverse model types. We fill this gap.

\section{Background}
Large-scale artificial intelligence training tasks show significant energy inefficiencies. For example, the GPUs used to train GPT-3 (175 B parameters) consumed almost 1,300 MWh of electricity—about the monthly consumption of 1,450 U.S. homes~\cite{brown2020language}. About 1.65 billion GJ ($\approx 458$ TWh) in global data centers drew about 2\% of world electricity consumption~\cite{IEA2023}. Nevertheless, average GPU use within these facilities stays only 30–50\%~\cite{weng2022mlaas}, suggesting that a significant portion of power draw happens in idle or stalled states (e.g., synchronizing or waiting on memory).

Further amplifying this waste are cooling and infrastructure overheads. Typical data center Power Usage Effectiveness (PUE) values range from 1.1 to 1.6, thus for every watt used by IT equipment, an additional 0.1–0.6 W is consumed on cooling and power distribution~\cite{IEA2023}. Thus, ancillary energy costs magnify inefficiencies in IT use.

Another important energy sink is data movement across nodes as well as on-chip. Since accessing HBM or DRAM often consumes an order of magnitude more energy per bit than on-chip compute~\cite{dao2022flashattention}, memory stalls and off-chip transfers can rule execution time in modern deep networks. Training workloads thus spend a large amount of runtime stalled on data I/O, so further lowering effective performance-per-watt.

These findings—low hardware use, high cooling overhead, and costly data movement—emphasize the need of \emph{eco-driven co-design} methods that jointly optimize hardware and software to maximize useful computation per joule.

\subsection{Industry Case Studies and Practitioner Insights}
Meta greatly expanded its AI infrastructure by adding two new clusters, each with 24,576 NVIDIA H100 GPUs for training Llama 3. This resulted in amazing throughput-per-watt thanks to custom network fabrics and open hardware designs\cite{Lee2024MetaGenAI}. NVIDIA released the improved H200 GPU, which has better memory bandwidth (4.8 TB/s) and efficiency. It can do almost twice as many LLM inferences as the H100\cite{CRN2023H200,Modular2025H100A100}. AMD's MI300X GPUs became very popular, especially with Meta and Microsoft using them for large-model inference workloads. They showed up to 1.9 times better performance per watt than earlier generations\cite{AMD2023MI300Xlaunch}. AWS used Trainium2 at an unprecedented scale—up to 400,000 chips—for training large models, which saved a lot of power per operation\cite{SiliconAngle2025AWSChips,AWS2024Trainium2}. Google Cloud moved forward with TPU v5e and gave a sneak peek at TPU v6 ("Trillium"), which made each training step 67\% more energy efficient\cite{GCP2023TPUv5e,Google2024Trillium}.

\subsection{Energy–Performance Tradeoffs}

Historically, AI research prioritized model accuracy and time-to-train over energy efficiency, leading to ever-increasing model sizes and computational demands. Only recently has the community begun to emphasize \emph{energy-conscious training}.  For example, power-capping experiments at a supercomputing center showed that throttling GPU power to 90\% of peak reduced overall energy consumption by 12–15\% while increasing training time by merely 3\%~\cite{dong2022slurmpowercap}.  Similarly, techniques such as early stopping and more effective hyperparameter search can prevent wasted computations with minimal impact on final model accuracy~\cite{strubell2019energy}.  

These findings demonstrate that significant efficiency gains are achievable without sacrificing performance, simply by eliminating present inefficiencies.  Yet the current landscape remains wasteful: infrastructure overhead (PUE\textgreater1.1), excessive data movement, and idle hardware contribute to large energy losses~\cite{henderson2020towards, IEA2023}.  Even routine model training can be made far greener, though large-scale runs highlight the extreme end: for instance, training GPT-3 consumed approximately 1\,300\,MWh—enough to power over 1\,400 U.S. homes for a month~\cite{brown2020language}.  

In the sections that follow, we explore hardware and software-level optimizations that directly target these inefficiencies to achieve more computation per joule—i.e., higher performance-per-watt.

\section{Optimization Techniques}
Research on two levels has been conducted to increase the sustainability of artificial intelligence: (1) \textbf{Hardware-level innovations}—designing and using computing hardware that delivers more performance per watt; and (2) \textbf{Software-level optimizations}—developing algorithms and systems that use hardware resources more efficiently. These two levels are complimentary; co-designing hardware and software for efficiency yields the best results. We go through each in turn, noting industry developments from 2021–2025 and representative studies that measure the gains.

\subsection{ Hardware-Level Optimizations}

\subsubsection{ Accelerators for Specialized Artificial Intelligence}
AI chips targeted at optimizing throughput per watt~\cite{reagen2021mlsurvey} have evolved rapidly over the past few years. While companies like Google have developed Tensor Processing Units (TPUs~\cite{jouppi2023tpuv4}, GPUs from NVIDIA and AMD have added dedicated tensor processing cores for deep learning~\cite{NVIDIAA100, AMDMI250X}. Others (e.g., Graphcore IPUs, Cerebras WSE) introduced custom accelerators~\cite{reagen2021mlsurvey}. Moving away from general-purpose designs to ASICs tailored for matrix math and parallelism~\cite{reagen2021mlsurvey} is a recurring motif.

By 2022, Google's TPU v4 and NVIDIA's third-generation Tensor Core GPUs (A100) embodied state-of-the-art. Mixed-precision matrix units included in these accelerators can run INT8 operations or FP16/BF16 at surprisingly higher efficiency than FP32 units~\cite{micikevicius2018mixed}.

With a \textbf{400 W} power envelope, the NVIDIA A100, for instance, achieves far better perf/W by delivering up to \textbf{312 TFLOPS} of BF16 throughput. Google's TPU v4 uses systolic arrays and advanced cooling, allowing it to reach high performance with efficiency— a recent study reported that TPU v4 is \textbf{1.2–1.7×} slower than A100 while using \textbf{1.3–1.9×} less power for the same tasks~\cite{jouppi2023tpuv4}. This corresponds to almost a \textbf{2×} improvement in performance-per-watt (e.g., 1.7× the speed at ~0.7× the power)~\cite{jouppi2023tpuv4}.

\begin{figure}[ht]
  \centering
  \includegraphics[width=0.9\linewidth]{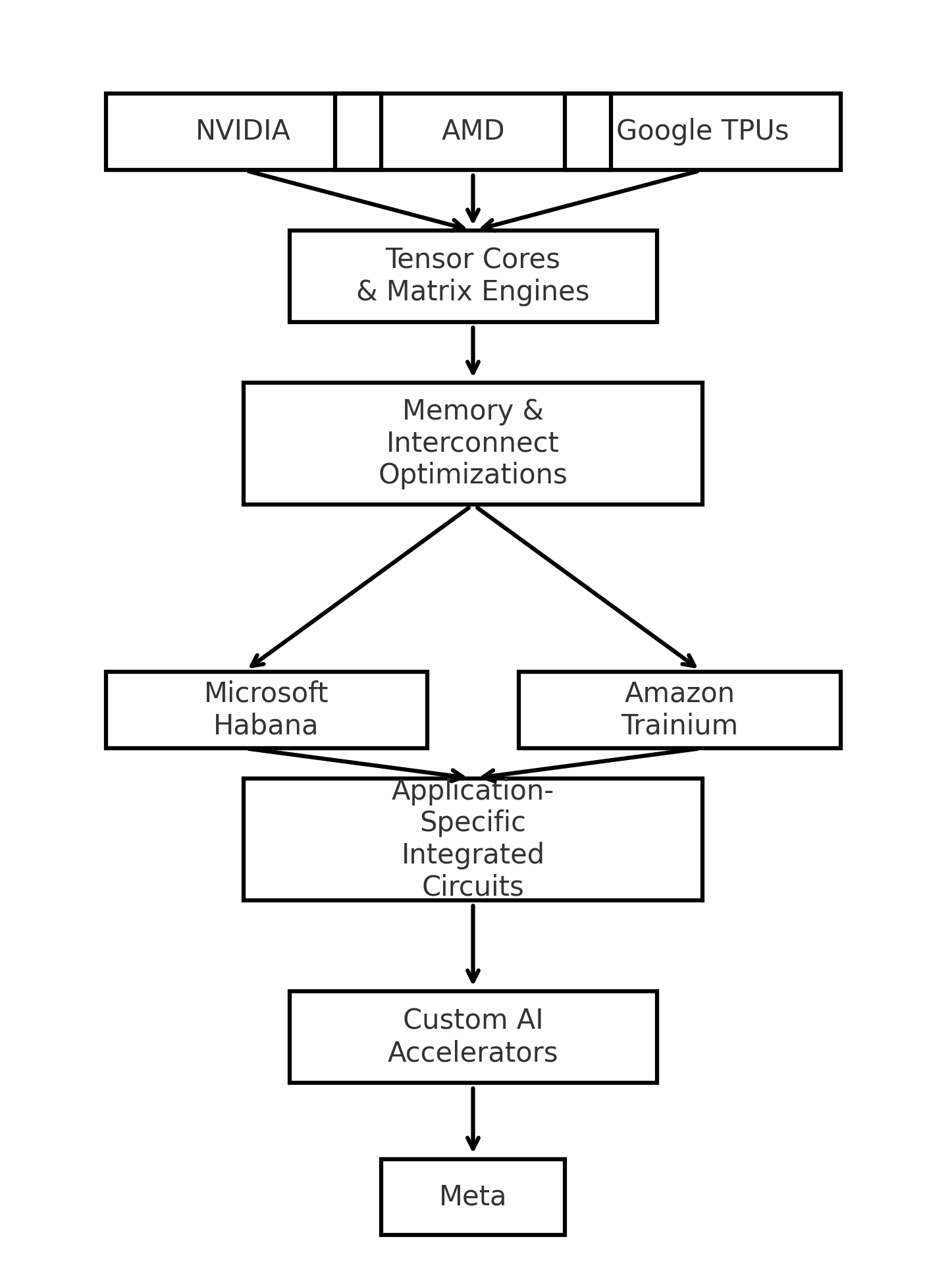}
  \caption{Hardware-level developments for training in sustainable artificial intelligence. To maximize efficiency, major vendors (NVIDIA, AMD, Google TPUs) have included specialized compute units (Tensor Cores, matrix engines) and high-bandwidth memory/interconnect optimizations. Targeting particular workloads, new competitors, including Microsoft (Habana) and Amazon (Trainium), have developed custom artificial intelligence chips (ASICs). Boosting performance-per-watt in model training depends critically on these domain-specific accelerators and architectural improvements.}
  \label{fig:hw_innovations}
\end{figure}

\subsubsection{ Architectural Innovations} 
Several architectural changes within contemporary artificial intelligence accelerators increase energy efficiency. One is the inclusion of matrix multiply engines (e.g., Tensor Cores~\cite{NVIDIAA100}, Google TPU MXUs~\cite{jouppi2023tpuv4}) that perform dense multiplication-add operations with lower precision and massive parallelism – so greatly increasing throughput per unit energy. Additionally used are optimized memory hierarchies and high-bandwidth memory (HBM). In 2022, flagship GPUs like NVIDIA A100/H100 and AMD MI250X featured HBM2e or HBM3 memory stacks delivering $>2\,$TB/s bandwidth~\cite{sheaffer2022h100, AMDMI250X}, so lowering the energy cost of moving data from memory into compute units. Fast interconnects (NVLink~\cite{nvidia2023nvlink}, InfiniBand~\cite{mellanox2021infiniband}, PCIe Gen5~\cite{pcisig2019pcie5}, and optical links) have also been used to speed up multi-accelerator communication – vital for distributed training efficiency so that power isn't wasted during synchronisation. To dynamically reconfigure network topology, Google's TPU v4 included optical circuit switches (OCS), so improving scaling efficiency and lowering of communication overhead~\cite{jouppi2023tpuv4}. This type of innovation guarantees that, when several chips cooperate, they do so with minimum idle time and least extra energy for data exchange. To increase perf/W in model training, domain-specific accelerators and architectural enhancements are essential Figure~\ref{fig:hw_innovations}

\subsubsection{ Efficiency Metrics of New Hardware}

Benchmark results clearly show the effects of these hardware-level improvements of efficiency. Not only does NVIDIA's most recent Hopper-generation GPU, the H100 (2022), boost raw speed, but it also claims 3× higher performance-per-watt than the previous A100 generation~\cite{sheaffer2022h100}. This results from 4 nm process technology (more energy-efficient transistors), more Tensor Cores that support new low-precision formats (FP8), and power optimizations like dynamic voltage–frequency scaling~\cite{sheaffer2022h100}. Similarly, AMD's Instinct MI250X GPU (7 nm, launched 2021) introduced a chiplet design with 2 GPU dies and stacked HBM2e memory, excelling in HPC efficiency—the Frontier supercomputer employing MI250X achieved 62.68 GFLOPS/W, topping the Green500 energy-efficient rankings in 2022~\cite{ TOP500}. With NVIDIA A100 nodes, this was over 2× the efficiency of previous-generation GPU systems (\(\sim26\) GFLOPS/W was state-of-the-art ~\cite{TOP500}.

Custom artificial intelligence chips from cloud providers have also surfaced outside of GPUs and TPUs. With an emphasis on high perf-per-dollar and perf-per-watt, Amazon's Trainium (2021) is an ASIC for AWS cloud that provides competitive training performance~\cite{aws2021trainium}. AWS claims Trainium instances deliver 2× higher throughput at 40\% lower cost than equivalently powered GPU instances). Another such Intel's Habana Gaudi2 (2022) is: built on 7 nm, Gaudi2 maintains somewhat better efficiency while outperforming NVIDIA A100 in throughput on some models using 24 tensor cores and 96 GB HBM. With Gaudi2's 600 W TDP against 400 W on A100, Intel reported 1.9× higher ResNet-50 training speed than A100—roughly 1.3× the performance-per-watt in that scenario~\cite{shilov2022gaudi}. These varied hardware choices—Graphcore IPUs, Cerebras WSE's wafer-scale engine, etc.—show a trend toward domain-specific architecture: by matching chips to matrix operations and deep learning patterns, they achieve more work with less energy than general-purpose devices.

All told, hardware-level optimizations—from completely new artificial accelerators to specialized functional units and memory systems—have produced notable efficiency gains. The best systems show 2–3× improvements in training performance per watt over 2–3 year old designs~\cite{sheaffer2022h100, TOP500}. Expecting to follow this trend are the next generation (e.g., NVIDIA's intended Blackwell GPUs, Google TPU v5, and AMD MI300 with 3D stacking). The basis is hardware, but, as we will discuss next, fully realizing energy savings also depends on smart software to keep these chips running in their ideal condition.

\subsubsection{ Real-World Impact}

The H200 GPU from NVIDIA came out in 2025. It had a big improvement in memory speed and efficiency, which made training large-scale LLMs much faster and easier. Meta used 350,000 H100 GPUs and a fully co-designed network fabric to get rid of congestion, which made training on Llama 3 very efficient\cite{Lee2024MetaGenAI}. Microsoft Azure and Oracle, two of the biggest cloud providers, widely used AMD's MI300X GPUs because they offered significant performance-per-watt improvements (up to 1.9$\times$ over MI250X)\cite{AMD2024AdvAI}. AWS showed that Trainium2 could work well on a large scale by making huge clusters for Anthropic's training needs and saving a lot of power compared to GPUs\cite{SiliconAngle2025AWSChips,AWS2024Trainium2}. Google's TPU v5e and upcoming TPU v6 made a big difference in how much energy they used, cutting it by 67\% per training step. This shows that they are using the best scaling and sustainability practices in the business\cite{GCP2023TPUv5e,Google2024Trillium}.

\subsection{Software-Level Optimizations}

\begin{figure}[ht]
  \centering
  \includegraphics[width=0.9\linewidth]{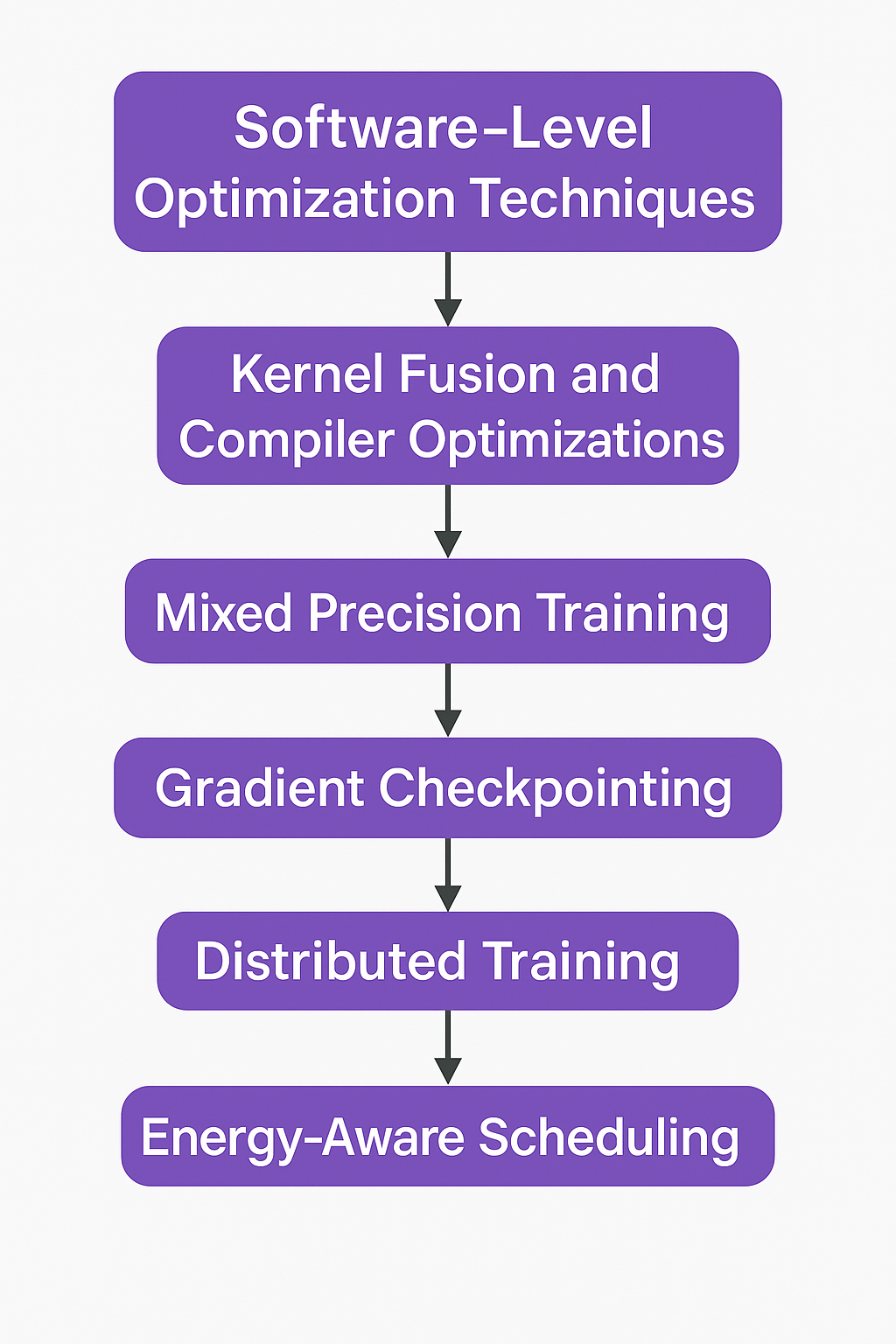}
  \caption{Software-level optimization techniques for energy-efficient AI training: (1) Kernel fusion and compiler optimizations—merging and optimizing compute kernels to reduce overhead~\cite{tillet2021triton}; (2) Mixed precision training—using lower numerical precision to speed up compute and save power~\cite{micikevicius2018mixed}; (3) Gradient checkpointing—saving memory by recomputing intermediate results, enabling larger batch sizes within the same memory budget~\cite{chen2016training}; (4) Distributed training—efficient parallelization across multiple devices for \textgreater{}90\% scaling efficiency~\cite{mattson2020mlperf}; (5) Energy-aware scheduling—job scheduling and power management strategies to minimize energy waste~\cite{dong2022slurmpowercap}.}
  \label{fig:sw_opt}
\end{figure}

From the software standpoint, many approaches can greatly increase training's energy efficiency without calling for new hardware. These improvements guarantee efficient use of the current computing capacity and elimination of pointless computation. Key strategies are mixed-precision training~\cite{micikevicius2018mixed}, kernel fusion~\cite{tillet2021triton}, memory optimization (e.g., checkpointing)~\cite{chen2016training}, improved scheduling~\cite{dong2022slurmpowercap}, and scalable distributed training algorithms~\cite{mattson2020mlperf} (see Figure~\ref{fig:sw_opt}). We evaluate every one of these and how they affect performance-per-watt.

\subsubsection{Mixed Precision Training (AMP)}

Adoption of Automatic Mixed Precision (AMP) techniques~\cite{micikevicius2018mixed} has been among the most powerful software optimizations. Many networks can be trained using 16-bit floats (FP16) for most operations, with minimum loss in model accuracy~\cite{micikevicius2018mixed}, instead of performing all computations in 32-bit floating point (FP32). Lowering precision helps specialized hardware units (Tensor Cores, etc.) to run more operations in parallel, so accelerating training by reducing memory bandwidth needs. Dramatic increases are shown by empirical results: in many cases~\cite{micikevicius2018mixed}, mixed-precision in training can increase throughput by 2–3×. Switching from FP32 to AMP, for instance, often doubles performance (and roughly doubles performance-per-watt) on NVIDIA's GPUs since the GPU can use its tensor cores and achieves better cache utilization~\cite{micikevicius2018mixed,nvidia2023fp8}. The NVIDIA Hopper H100 goes further by including FP8 support—early experiments show that FP8 training can further improve speed and efficiency, maybe yielding 3–4× perf/W gains over FP32 on the same hardware~\cite{nvidia2023fp8}. Software frameworks like PyTorch and TensorFlow now include built-in AMP support (e.g., \texttt{torch.cuda.amp} module), so facilitating researchers' application of mixed precision. Using suitable precision—a simply software-driven improvement—one can significantly lower energy per training step for the same hardware and model.

\subsubsection{Kernel Fusion and Compiler Optimizations}

Contemporary deep networks run as a series of several small operations (layers, activations, etc.). Should every operation start independently on the hardware, there is overheads—that is, kernel launch times—that waste time and energy by means of memory reads and writes for intermediary results.
Kernel fusion solves this by aggregating several operations into one GPU kernel such that data stays on-chip and intermediate computations are effectively reusing~\cite{ginsburg2020gpu}. Elementwise operations or even whole subgraphs of computations can be automatically fused by compiler and graph optimizer XLA~\cite{xla2023overview}, TorchScript and JAX, NVIDIA's CUTLASS, and Triton compiler~\cite{tillet2021triton}. Eliminating extraneous memory traffic and launch overhead helps fused kernels to accelerate training throughput by \textbf{10–30\%} in real-world models~\cite{tillet2021triton}. One concrete example is Google's XLA, which, when turned on, can provide a \textbf{1.1×–1.3×} performance boost for different models~\cite{xla2023overview}. Likewise, the OpenAI Triton compiler is shown to significantly increase efficiency for particular operations~\cite{tillet2021triton} and lets custom fusion of PyTorch kernels possible. Generally, compiler-level optimizations guarantee that the computational units of the hardware are fed with work steadily, so lowering the relative overhead (and energy cost) of every training iteration.

\subsubsection{Memory Optimization and Gradient Checkpointing}

Throughout training, GPU memory is a limited and valuable resource. Many models cannot fit high-density data in memory or large batch sizes, thus either out-of-memory errors or small batch usage—which can underutilize the compute units, so wasting possible performance. Instead of storing all intermediate activations for backpropagation, one stores only a subset (checkpoints) and recomputes the others on-demand during the backward pass~\cite{chen2016training,zhang2022zeroinf}. Gradient checkpointing, sometimes known as rematerialization, trades extra compute for lower memory use. Often by 50\% or more, this can drastically reduce memory needs, allowing bigger batch sizes or models on the same hardware~\cite{zhang2022zeroinf}. Larger batches reduce the recomputation overhead by improving hardware use and throughput, so enabling more training work completed per unit time. Studies have shown that generally speaking, the compute overhead of checkpointing—due to recomputation—is typically far smaller than the gains from being able to use a larger batch size or avoid gradient accumulation steps, so reducing overall training time and energy. If checkpointing allows doubling the batch size, for example, the GPU can be kept \(\sim100\%\) busy (full utilization) instead of, say, 50\% busy, leading to almost 2× throughput—far outweighing the \(\sim5\%-20\%\) extra compute from recomputation~\cite{zhang2022zeroinf}. Many recent large models save memory by using frameworks like PyTorch, which in practice offer simple checkpointing APIs (e.g., \texttt{torch.utils.checkpoint})~\cite{pytorch2023checkpoint}. Software can thus unlock better perf/W by improving memory reuse and allocation: the model performs the same number of arithmetic operations, but finishes faster (hence less total energy) because it used the hardware more effectively with the available memory.

\subsubsection{Distributed Training Efficiency}

Modern training models sometimes demand splitting of work among several GPUs/TPUs. Scaling distributed training effectively is difficult, though; if poorly controlled, adding more chips can result in declining returns and a lot of wasted energy on powered but idle devices. High-efficiency parallel training has lately attracted attention in engineering and research. Techniques include gradient compression~\cite{lin2018deep}, model parallelism (sharding the model layers across devices)~\cite{huang2019gpipe}, and pipeline parallelism (streaming micro-batches across different model stages)~\cite{narayanan2019pipedream}. The secret is overlap and load balancing—that is, overlapping communication with computation to guarantee each accelerator has a roughly equal amount of work. Industry benchmarks show almost linear scaling in well tuned systems: Using $\geq 1000$ GPUs can help to achieve $>90\text{–}95\%$ scaling efficiency for large models using optimal communication libraries and topology-aware algorithms~\cite{mlperf2023}. Google claimed that their most recent TPUv5 pods, on a 175B-parameter GPT-3 model across thousands of TPU chips, attained $99.9\%$ scaling efficiency~\cite{google2023tpuv5}, meaning almost perfect linear speedup, which implies each chip’s resources were effectively used. Likewise, from 256 to 384 accelerators Habana’s Gaudi2 showed $\sim95\%$ scaling efficiency~\cite{habana2022scaling}. 
For energy, effective scaling is important since an N-node job that achieves near-linear scaling will train the model N times faster but consume only somewhat more total energy than a 1-node job (the decreased time balances the increased energy from extra nodes). Conversely, ineffective scaling—that is, 50\% efficiency—means half the added compute power is wasted—those GPUs still draw power but do nothing half the time. Advances in distributed training software (like Facebook’s FairScale~\cite{fairscale2021} and PyTorch FSDP~\cite{pytorch2021fsdp}, DeepSpeed ZeRO~\cite{deepspeed2021}, and NVIDIA’s NCCL library~\cite{nccl2022} for fast all-reduce) have been vital to ensure large-scale training is energy-proportional; you roughly get what you pay for in terms of extra power.

\subsubsection{Scheduling and Energy Awareness}

Another developing tool for software optimization is at the system orchestration level, which raises awareness of energy issues for training jobs. Runtime application of techniques including dynamic frequency scaling and GPU power capping can help to remove inefficiencies. For a minor performance loss~\cite{dong2022slurmpowercap}, capping GPU power limit to a somewhat lower level can, as noted in the Background, produce double‐digit percentage energy savings. Programmatically controlling these caps or changing clocks is made possible by software tooling including AMD's ROCm SMI~\cite{amd2022rocm_smi} and NVIDIA's DCGM~\cite{nvidia2022dcgm}. Furthermore being improved for energy-aware scheduling are job schedulers found in data centers (such as Slurm, Kubernetes). By including power monitoring and capping into Slurm, MIT Lincoln Lab researchers were able to schedule jobs in ways that lower peak power demand and exploit cooler periods~\cite{gadepally2022power}. Non-urgent training jobs can be run at night or when renewable energy supply is high, for example; jobs can be queued so that total power stays within an efficient range to avoid straining cooling systems~\cite{gadepally2022power}. Research on auto-scaling and instance right-sizing for cloud-based training—automatically adjusting resources to maximize efficiency (e.g., underutilized I/O-bound jobs might use smaller instances)—~\cite{kubernetes2022autoscaler}. Early results are encouraging even if these system-level approaches are still under development. Using holistic scheduling techniques helps one to keep GPUs running in ideal conditions (avoidance of low-utilization situations) and smooth out usage spikes, so lowering auxiliary energy costs (like cooling). Such policies combined with lower-level optimizations guarantees that, from algorithm to silicon, energy economy is taken into account at every level.

By conservatively combining the following independent gains:
\begin{itemize}
  \item Mixed precision training: $2\times$~\cite{micikevicius2018mixed}
  \item Kernel fusion (compiler optimizations): $1.2\times$~\cite{tillet2021triton}
  \item Gradient checkpointing (larger batch sizes): $1.5\times$~\cite{chen2016training}
  \item Power capping (energy‐aware scheduling): $1.12\times$~\cite{dong2022slurmpowercap}
\end{itemize}
the combined performance-per-watt uplift is
\[
2.0 \times 1.2 \times 1.5 \times 1.12 \approx 4.0,
\]
showing that fully embracing software-level optimizations can deliver well over a $3\times$ improvement in training energy efficiency. Mixed precision by itself usually provides a multi-fold boost~\cite{micikevicius2018mixed}, compiler and memory optimizations add further gains~\cite{tillet2021triton,zhang2022zeroinf}, and efficient distributed algorithms prevent waste at scale~\cite{mlperf2023,dong2022slurmpowercap}. Most importantly, these methods are essentially complimentary: state-of-the-art training runs (such as those in MLPerf or corporate internal benchmarks) now routinely employ all of the above: e.g., using AMP, fused kernels, huge batch with checkpointing, and advanced optimizers on massive GPU clusters, plus careful scheduling. We can thus train models greener and faster than in past years. In the following part, we turn our attention to the quantification of such achievements: the instruments and measures applied to evaluate performance and energy in artificial intelligence training.

\subsection{Industry Adoption and Software Co-design}
\subsubsection{Meta's Llama 3 Clusters}
Meta has added two new clusters to its AI infrastructure. Each one has 24,576 NVIDIA H100 GPUs that are used to train Llama 3, which is built on Meta's open \textit{Grand Teton} platform\cite{Lee2024MetaGenAI,Wodecki2024LlamaCluster}. Using OCP Open Rack standards and PyTorch frameworks, this design can work with models like Llama 3 and newer. Meta wanted 350,000 H100 GPUs in data centers by the end of 2024. This is about the same as 600,000 H100s of compute power using clustering techniques\cite{Lee2024MetaGenAI}. The clusters use two 400 Gb/s fabrics (RoCE Ethernet and InfiniBand) to get rid of network bottlenecks. In fact, the RoCE-based cluster can run Llama 3 training at scale without any problems. Meta says that a 405 B-parameter Llama 3.1 model was trained on a 16 k-GPU H100 cluster with about 90\% of its power used up in 54 days. This shows how important it is to manage power and reliability at very large scales\cite{Lee2024MetaGenAI}. Meta's open design and use of the Grand Teton chassis, which combines power, cooling, and fabric, shows how important it is to design hardware and software together for long-term AI training\cite{Lee2024MetaGenAI}.

\subsubsection{NVIDIA H100/H200 Performance and Efficiency}
The NVIDIA H100 Hopper GPU (2022) is now the standard for efficient training, with up to 60 TFLOPS FP32 (2× A100) and 3 TB/s memory bandwidth\cite{Modular2025H100A100}. The H100 is very important because it has about three times the performance per watt of the A100 that came before it\cite{Modular2025H100A100}. The 4th-gen Tensor Cores, Hopper architecture improvements, and better cooling all contribute to this increase in efficiency. NVIDIA released the H200 in 2025. It was an improved version of the Hopper with 141 GB HBM3e at 4.8 TB/s—1.4 times the memory capacity and 1.4 times the bandwidth of the H100\cite{CRN2023H200}. The H200 is optimized for memory-bound workloads and has an LLM inference throughput that is about 1.6 to 1.9 times higher than the H100, for example, +90\% on Llama2-70B. It is compatible with H100 HGX systems, which makes upgrades easy\cite{CRN2023H200}. NVIDIA stresses energy-efficient performance even more: a server with eight H200 GPUs can do 32 PFLOPS of FP8 compute with 1.1 TB/s of total memory. NVIDIA's engineering leads say that the company is ``relentlessly pursuing energy-efficient performance'' through both hardware and software co-design.

\subsubsection{Deployments of the AMD Instinct MI300X}
The AMD Instinct MI300 series (CDNA 3) became popular in 2024–2025 as a training accelerator with a lot of memory and good performance. The MI300X GPU has 192 GB of HBM3 and can transfer data at a rate of about 5.3 TB/s. This is very useful for big models. A single MI300X has about 40\% more cores, 1.5 times more memory, and 1.7 times more bandwidth than a MI250X\cite{AMD2023MI300Xlaunch}. AMD's MI300A (GPU+CPU APU variant) gets about 1.9 times better FP32 performance-per-watt on AI/HPC workloads than the MI250X. This is because it has 3D-stacked integration and unified CPU-GPU memory\cite{AMD2023MI300Xlaunch}. This helped AMD go above and beyond its ``30×25'' efficiency goal, giving it a 38× node-level energy efficiency improvement from 2020 to 2025. Meta's use of MI300X for Llama inference in production and Microsoft's use of MI300X accelerators on Azure are examples of real-world adoption. Oracle Cloud is also planning bare-metal MI300X instances with clusters of up to 131,072 GPUs\cite{AMD2024AdvAI}. Early tests show that an 8× MI300X OAM server can do LLM inference throughput about 1.6 times better than an 8× H100 HGX system (BLOOM-176B). It can also host a 70 B-parameter model entirely in memory\cite{AMD2023MI300Xlaunch}. According to reports, AMD's mid-2025 MI350X/MI355X (CDNA 4) cards will have 3–4 times the training performance of the MI300X and up to 35 times the inference throughput, showing that efficiency is still improving\cite{Tomshardware2025MI350X}.

\subsubsection{AWS Trainium2 and Custom Silicon}
Amazon Web Services has put more money into making its own AI chips to make training more efficient on a large scale. The AWS Trainium2 (2024) delivers 1.29 PFLOPS of FP8 per chip, which is up to four times more than its predecessor. It powers EC2 Trn2 instances, which have 16 chips, 20.8 PFLOPS FP8, 1.5 TB HBM3, and 3.2 Tb/s networking. AWS says that Trn2 is 30–40\% better at price-performance than similar GPU P5 instances and uses less power\cite{AWS2024Trainium2}. In early 2025, AWS showed off a 400,000-chip Trainium cluster for training Anthropic's model. It was a big deal because it showed how scalable and energy-efficient it was. AWS also released Trn2 UltraClusters, which are nodes of 64 chips linked by NeuronLink optical interconnect for high parallel efficiency. They also announced Trainium3, which is expected to double performance and improve energy efficiency by about 50

\subsubsection{Google Cloud TPU v5e and TPU v6, also known as ``Trillium''}
Google's TPU v5e (GA Nov 2023) has a training price-performance that is 2.3 times better than TPU v4 on MLPerf benchmarks. This is because of improvements to the architecture and the use of INT8+BF16 mixed-precision\cite{GCP2023TPUv5e}. Google showed a 50,000-chip TPU v5e Multislice LLM training run at 53\% parallel efficiency, which is one of the biggest AI jobs ever\cite{GCP2023TPUv5e}. Google showed off TPU v6 (``Trillium'') in October 2024. It had 4 times better training performance and used 67\% less energy per training step\cite{Google2024Trillium}. Trillium also doubles the memory and interconnect bandwidth compared to v5e, which cuts down on energy-intensive sharding. It can also scale up to 91 exaFLOPS on Jupiter fabric interconnects, which is almost linear.

\subsubsection{Apple Silicon and Efficiency on Devices}
Apple isn't a cloud provider, but its M3 Ultra SoC (Mar 2025) shows how well hardware and software can work together. It has a 32-core CPU, an 80-core GPU, and a 32-core Neural Engine. It has double the throughput of the M2 Ultra at the best power efficiency in the industry\cite{Apple2025M3Ultra}. UltraFusion die-to-die interconnect (2.5 TB/s) connects two dies without adding any extra energy use for multiple GPUs. Apple says that the Mac Studio with M3 Ultra can run a 600 B-parameter LLM locally (192 GB unified memory, dedicated ML accelerators), which shows that large-model inference can be done on a device with very little power use\cite{Apple2025M3Ultra}. This shows that optimizing at the SoC level—getting the most out of local memory and moving the least amount of data—gives you great performance per watt, which means AI can be used in more places than just data centers.

Table~\ref{tab:practitioner-checklist} is a useful summary of the most important best practices and proven benchmarks.

\begin{table*}[htbp]
  \caption{Checklist for Practitioners for Sustainable AI Training}
  \label{tab:practitioner-checklist}
  \begin{center}
    \begin{tabular}{|p{3cm}|p{13.5cm}|}
      \hline
      \textbf{Aspect} & \textbf{Best Practices and Benchmarks} \\
      \hline
      Hardware Selection &
      Select the newest accelerators, such as the NVIDIA H200 (up to 1.9× more inference throughput than the H100)~\cite{CRN2023H200}, the AMD MI300X (1.9× more performance per watt than the MI250X)~\cite{AMD2023MI300Xlaunch}, or the TPU v6 (67\% more efficient than the TPU v5e)~\cite{Google2024Trillium}. \\
      \hline
      Low-Precision Compute &
      Use AMP or INT8 training methods (like Google's TPU v5e with AQT quantization) to get up to 2.3× better price-performance~\cite{GCP2023TPUv5e}. \\
      \hline
      Memory Efficiency &
      Use high-bandwidth-memory GPUs like the NVIDIA H200 (4.8 TB/s HBM3e) or the AMD MI300X (192 GB HBM3) to cut down on energy-heavy memory accesses~\cite{CRN2023H200,AMD2023MI300Xlaunch}. \\
      \hline
      Custom AI Chips &
      Prefer cloud-specific chips like AWS Trainium2 (up to 4× the throughput of Trainium1 and 30-40\% better price-performance than GPU instances)~\cite{AWS2024Trainium2}. \\
      \hline
      Cluster Co-Design &
      Optimize networking (Meta’s RoCE/InfiniBand, Google’s TPU Multislice fabric) to keep scaling efficiency high (Meta: ~90\%; Google: \textgreater{}50\% at 50000 chips)~\cite{Lee2024MetaGenAI,GCP2023TPUv5e}. \\
      \hline
      Cooling Optimization &
      Use advanced cooling (e.g.\ liquid cooling for AMD MI355X) to avoid thermal throttling in dense racks~\cite{Tomshardware2025MI350X}. \\
      \hline
      Carbon Footprint &
      Choose regions with abundant renewables (e.g.\ Google’s carbon-free zones) to cut training CO$_2$ by up to 50\%~\cite{Google2024Trillium,CRN2024AWSChips}. \\
      \hline
    \end{tabular}
  \end{center}
\end{table*}

\section{Measurement and Benchmarking}
Driving development in energy-efficient artificial intelligence requires consistent measurement of performance and power consumption. The main metrics used for energy-aware evaluation are described in this part together with the tools and approaches usually used to benchmark AI training efficiency.

\subsection{Energy‐Efficiency Metrics}

FLOPS per Watt (FLOPS/W)—how many floating-point operations a system can perform for each watt of power consumed—is a fundamental metric~\cite{dongarra2003linpack}. Useful for hardware specification comparisons, FLOPS/W can be measured for sustained performance on a given workload or for peak theoretical performance. For instance, GFLOPS/W attained on the HPC LINPACK test is used as ranking metric~\cite{TOP500} in the Green500 list, an energy-oriented subset of Top500 supercomputers. As a task-specific metric in artificial intelligence training, one could evaluate training images per second per watt or training samples per joule.  

MLPerfTM Power is another crucial statistic that lately has been introduced. Beginning about 2022, the MLPerf Training benchmark suite included an optional power measuring track where submitters run standardized training tasks and log the energy consumed~\cite{mlperf2023}. For every submission, this enables reporting of metrics like ``time-to-train at X watts'' or conversely ``throughput per watt''. Though our emphasis here is on the technical energy aspect, a related idea is performance-per- dollar or perf-per-Watt-per-\$ ~\cite{strubell2019energy}, Whatever the particular metric, the objective is to normalize performance by power to evaluate systems on energy economy instead of raw speed by itself.

\subsection{Power and Energy Monitoring}

Measuring power use during training can be accomplished with either built-in sensors or outside equipment. Most contemporary CPUs and accelerators feature on-board power monitors reachable via APIs. For GPUs (exposed via the \texttt{nvidia-smi} tool), NVIDIA's NVML (NVML) for instance, offers real-time power draw readings~\cite{NVIDIA2021NVML}. AMD offers a comparable interface for Radeon/Instinct GPUs~\cite{amd2022rocm_smi} using \texttt{rocm-smi}. These devices track power (in watts) at, say, 1-second intervals while a training job runs, so generating a trace of energy use. Studies have validated these readings against external power meters and given insight into their accuracy—usually, the internal sensors are sufficiently accurate (within a few percent) for most comparison purposes~\cite{li2021gpu_measure}, though external high-precision meters are used for rigorous benchmarking~\cite{rios2023pdu}. Apart from tools particular to vendors, frameworks such as PyTorch-Lightning or experiment trackers can combine power logging~\cite{Falcon2021Lightning}. Though those are more pertinent for inference on CPU, there are also platform-agnostic tools including Intel Power Gadget (for CPU power)~\cite{Intel2022PowerGadget} and Linux's RAPL interface for system power~\cite{jain2021rapl}. Some research systems~\cite{rios2023pdu} use external metered PDUs (power distribution units) or wall-socket power meters for end-to-end tracking (including whole server or cluster power).

\subsection{Carbon Footprint Tools}

Another measure is translating energy consumption into carbon emissions using tools for carbon footprints. Popular among tools like Carbontracker (introduced in late 2019)~\cite{lacoste2019carbontracker} and more recent libraries like CodeCarbon~\cite{codecarbon2021} are Often based on location and time, these tools estimate CO$_2$ emissions of training by combining logged energy consumption with the carbon intensity of the electricity grid being used (~\cite{lacoste2019carbontracker}. To indicate that ``training this model emitted X kg of CO$_2$'', for instance, CodeCarbon can take power readings and multiply by an emission factor (kg CO$_2$/kWh). Important for environmental impact reporting, these measurements have been applied in studies to evaluate the carbon footprint of several model training strategies~\cite{strubell2019energy}. Although our analysis concentrates on performance-per-watt, it is important to keep in mind that lowering CO$_2$ per training task is usually the ultimate aim and these tools enable to measure that higher-level metric.

\subsection{Benchmarking Workloads}

Standardized benchmarks are absolutely essential for fairly evaluating hardware or optimization methods. Organizations benchmark their hardware/software stacks, and MLCommons—the consortium behind MLPerf—offers a set of representative training tasks—image classification on ResNet, language modeling on BERT, recommendation, object detection, etc.—where organizations benchmark their hardware/software stacks~\cite{mlperf2023}. Not only time-to-training but also power traces can be recorded in MLPerf logs, so allowing computation of training energy~\cite{mlperf2023}. One submission might, for example, report 1 hour's training ResNet-50 with an average cluster power draw of 10 kW, so approximating 10 kWh consumed. Some entries have in fact included this data as of MLPerf v2.0+, allowing direct perf-per-watt comparisons between systems~\cite{mlperf2023}. Apart from MLPerf, academic researchers evaluate particular optimizations using tailored micro-benchmarks (e.g., training a small model for N steps and measuring joules).

\subsection{Tools for Developers}

There are several easily available tools for those trying to maximize their own training runs. By optimizing those, NVIDIA's DLProf and Nsight Systems can profile GPU use and point out inefficiencies (such as low SM occupancy or memory bottlenecks)~\cite{nvidia2022dlprof,nvidia2022nsight} indirectly helping to minimize energy waste. Regarding power, one can log it using \texttt{nvidia-smi dmon} and observe whether there are times when GPUs are consuming maximum power but usage is low (indicative of inefficiencies)~\cite{NVIDIA2021NVML}. Cluster managers are beginning to include energy metrics in multi-node configurations: e.g., if the hardware supports it, Slurm can record per-node energy usage; newer Slurm versions allow scheduling constraints like energy limits for a job~\cite{gadepally2022power,slurm2023energy}. These capabilities allow what MIT Lincoln Laboratory showed—enforce power caps and maximize scheduling to save energy without user involvement~\cite{gadepally2022power}. With a growing ecosystem of tools supporting it, measuring and benchmarking energy in AI training has become a mainstream activity in recent years from a niche concern.

\section{Research Gaps and Future Directions}
Although artificial intelligence hardware and software have advanced significantly, several important research questions still surround creating really energy-efficient training systems.

\subsection{Data Flow and the Memory Wall}
In AI systems, memory access still controls most energy consumption. Moving data from off-chip memory often costs 10–100$\times$ more than computation~\cite{horowitz2014computing, mittal2021survey}. HBM and big on-chip caches notwithstanding Understudied for mainstream GPU architectures~\cite{mittal2021survey} are processing-in-memory (PIM), near-memory computation, and dataflow reordering.

\subsection{Lack of Standardized Energy Benchmarks}
With few reports of power or energy measurements, most artificial intelligence benchmarks center on accuracy and throughput. Energy is an optional measurement track ~\cite{mlperf2023}; industry-wide adoption is lacking though. Standard measurements of joules-per-epoch and CO$_2$ emissions could direct more environmentally friendly growth~\cite{strubell2019energy}.

\subsection{Hardware-Software Co-Optimization Barriers}
Academic research sometimes looks at hardware or algorithms alone. Co-design frameworks that optimise AI models and compiler jointly for target hardware under energy constraints, including neural architecture search (NAS) techniques prioritising efficiency~\cite{cai2020once} and autotuning compiler that minimise energy rather than latency alone, are needed.

\subsection{Underutilization and Poor Energy Proportionality}
Many artificial intelligence training centers run far below maximum capacity, so wasting energy. Techniques including dynamic job packing and GPU sharing (MIG)~\cite{ NVIDIA2021Mig}, as well as adaptive workload scaling, demand more study to guarantee energy scales with compute demand~\cite{henderson2020towards}.

\subsection{Sustainable Hardware Lifecycle}
Manufacturing high-end GPUs and accelerators has non-trivial embodied carbon. Hardware reuse techniques, multi-generation model migration, and circular design concepts ought to be studied. As much as enhancing running efficiency~\cite{wu2023lca} extending the useful life of AI hardware can help sustainability.

\subsection{ Novel Compute Paradigms}
Massive energy reductions (10$\times$ or more) are possible with analog, photonic, and in-memory computing architectures; but, programmability, noise tolerance, and integration~\cite{chen2020analog,shastri2021photonics} suffer. A major research challenge still is bridging the software-hardware gap for various technologies~\cite{mittal2021survey}.

The road to sustainable artificial intelligence calls for rethinking compute, data access, software design, and evaluation criteria from the ground up in addition to maximizing today's hardware.

\section{Conclusion}
In this review, we have synthesized recent advances in \emph{eco-driven} performance optimization for AI training, spanning both hardware and software domains. We surveyed key inefficiencies—underutilized compute resources and excessive cooling overhead—and highlighted architectural innovations (tensor/matrix cores, high-bandwidth memory, efficient interconnects) that modern accelerators (GPUs, TPUs, custom ASICs) employ to boost performance-per-watt. On the software side, techniques such as mixed-precision training, kernel fusion, memory-efficient checkpointing, and scalable distributed algorithms all contribute multiplicative improvements, often combining to exceed a $\times3$ uplift in effective perf/W.

Our comparative analysis underscores that recent hardware such as NVIDIA’s H100 and Google’s TPU v4 achieve multiple times higher perf/W than their predecessors, validating the energy-centric design paradigm. However, critical research gaps remain: the \emph{memory wall} and data‐movement costs, the absence of standardized energy benchmarks beyond MLPerf Power, and barriers to hardware–software co‐optimization that could automate energy-aware neural architecture search and autotuning. Emerging evaluation tools—carbon trackers like Carbontracker and CodeCarbon—further extend metrics from wattage to CO\(_2\) emissions, linking technical efficiency to environmental impact.

Table~\ref{tab:practitioner-checklist} gives practitioners clear steps to take to use these environmentally friendly AI training methods.

Ultimately, achieving truly sustainable AI training demands holistic co‐design: from circuit‐level innovations to compiler optimizations, runtime schedulers, and policy frameworks. As the community embraces \emph{Green AI} principles and integrates energy metrics into mainstream benchmarks and tooling, we can realize both faster and greener training. We hope this review serves as a comprehensive reference and a call to action for researchers and practitioners to push the frontiers of eco-driven AI.

\bibliographystyle{IEEEtran}
\bibliography{references}

\end{document}